\documentclass[%
 reprint,
 amsmath,amssymb,
 aps,
 array,
 graphicx,
 esint,
]{revtex4-1}
\usepackage[T1]{fontenc}
\usepackage[latin9]{inputenc}
\usepackage{graphicx}
\usepackage{dcolumn}
\usepackage{bm}
\usepackage{color}

\providecommand{\tabularnewline}{\\}

\begin{document}

\preprint{APS}

\title{Effective viscosity of a suspension of flagellar beating microswimmers: Three-dimensional modeling}

\author{Levan Jibuti}
\author{Walter Zimmermann}
\email{walter.zimmermann@uni-bayreuth.de}
\affiliation{
 Theoretische Physik I, Universit�t Bayreuth, 95440 Bayreuth, Germany
}

\author{Salima Rafa\"i}
\author{Philippe Peyla}
\email{philippe.peyla@univ-grenoble-alpes.fr}%
\affiliation{
 LIPhy,Universit\'e Grenoble Alpes and CNRS, F-38402 Grenoble, France
}

\begin{abstract}
Micro-organisms usually can swim in their liquid environment by flagellar or ciliary beating.
In this numerical work, we analyze the influence of flagellar beating on the orbits
of a swimming cell in a shear flow. We also calculate the effect of the flagellar beating on the rheology of a dilute suspension of micro-swimmers.
A three-dimensional model is proposed for \textit{Chlamydomonas Reinhardtii} 
swimming with a breaststroke-like beating of two anterior flagella modeled by two counter-rotating fore beads. 
The active swimmer model reveals unusual angular orbits in a linear shear flow. 
Namely,  the swimmer sustains orientations transiently across the flow.
Such behavior is a result of the interplay between shear flow
and swimmer's periodic beating motion of flagella which exert internal torques on the cell body. This peculiar behavior has some significant consequences 
on the rheological properties of the suspension. We calculate the 
Einstein's viscosity of the suspension composed of such isolated modeled microswimmers 
(dilute case) in a shear flow. We use numerical simulations based on a Rotne-Prager like approximation for hydrodynamic interaction between simplified flagella and the cell body. 
The results show an increased intrinsic viscosity for active swimmer suspensions 
in comparison to non-active ones as well as a shear thinning behavior
in accordance with previous experimental measurements {[}Phys. Rev. Lett. {\bf 104}, 098102 (2010){]}.

\begin{description}

\item[PACS numbers]
47.63.Gd, 47.63.mf, 83.10.Pp

\end{description}
\end{abstract}

\pacs{Valid PACS appear here}
\maketitle

\section{Introduction}

Self-propelled particles and micro-organisms able to swim on a microscopic scale have attracted enormous
interest over the last few years \cite{Lauga:2009.1,KochDL:2011.1,Guasto:2012.1,Lauga:2016.1,Bechinger:2016.1}. Typical examples of microswimmers include 
biological organisms: microalgae, bacteria, and sperm cells as well as artificial 
swimmers \cite{Dreyfus:2005.2,Keaveny:2008.1,FischerT:2013.1,KochDL:2011.1,Guasto:2012.1,Lauga:2016.1,Bechinger:2016.1}. There are two major categories of swimmers: 
''pullers'' and ''pushers.'' A puller, like for example the 
micro-alga \textit{Chlamydomonas Reinhardtii} (\textit{CR}), has two anterior flagella that pull 
the fluid toward the cell body along the swimming direction while a pusher 
(like the bacteria \textit{Escherichia coli}) pushes the fluid behind the cell body opposite to 
the swimming direction \cite{Lauga:2009.1}. Designing controllable microswimmers, 
capable of detection \textit{in vivo} 
and carrying a drug to treat and target localized diseases, is one of the most desired objectives in biophysics. 
Such a microswimmer should employ special tactics to overcome low Reynolds number constraints 
for locomotion \cite{Purcell:1977.1} 
as well as control their swimming direction or cross-streamline migration \cite{Kaoui:2008.1} 
in the flow (e.g. Poiseuille flow in blood streams).

At a macro-scale, active fluids are suspensions of self-propelled  micro-organisms (or artificial particles)
which by moving, spinning or deforming, significantly alter the 
macroscopic properties of the fluid \cite{Haines:2008.1} like its effective viscosity. 
After the emergence of theoretical models and simulations of rheological properties 
\cite{Hatwalne2004,Saintillan:2010.2,Haines:2008.1,Haines2011,Heidenreich:2011.1,Cui_Z:2011.1,Ishikawa:2007.1}, 
an increasing number of experiments has been published on experimental measurements on effective viscosity 
of active suspensions \cite{Aranson:2009.1,Peyla:2010.1,Mussler:2013.1,Clement:2013.1,Clement:2015.1}. 
Sokolov and Aranson \cite{Aranson:2009.1} measured the viscosity of pusher type bacterial suspension. 
They report strong decrease of the effective viscosity (up to a factor of $7$). Lopez \textit{et al.} \cite{Clement:2015.1} 
even showed that for semi-dilute \textit{E. Coli}  suspensions the viscous resistance to shear can vanish. 
The viscosity of planktonic suspension of \textit{CR} (puller type)
has been measured \cite{Peyla:2010.1}
and a  significantly higher viscosity was found compared to suspensions containing the same volume fraction of dead cells. 
Suspensions of living
microalgae also show a shear-thinning behavior. In another experiment, Mussler \textit{et al.} \cite{Mussler:2013.1} confirmed 
the previous results using two different  geometries (Taylor-Couette and cone-plate) and show that gravity does not play a role in 
the enhancement of the viscosity unlike suspensions of bottom-heavy \textit{Chlamydomonas nivalis} 
\cite{Stocker:2009.1,Clement:2013.1,Ishikawa:2007.1}.

The physical interpretation of this peculiar rheology was first introduced by Hatwalne \cite{Hatwalne2004}. It can be summarized as follows. An elongated microswimmer (rod-like bacteria for example), 
once immersed in a simple shear flow of shear rate $\dot{\gamma}$, spends a long time compared to $\dot{\gamma}^{-1}$ in the extensional 
direction of the shear flow: this is known as Jeffery orbits \cite{JefferyGB:1922.1}. 
And since each microswimmer is modeled as a permanent force dipole, it increases for pushers or decreases for pullers 
the off-diagonal stress tensor, respectively, resulting in a decrease or an increase of the effective viscosity.
The majority of models relies on this anisotropic orientation distribution of microswimmers in the flow. Such an assumption is natural for the suspensions of pusher 
type bacteria that have a rod-like shape or for gravitactic swimmers that are oriented by  gravity \cite{Ishikawa:2007.1}. However, it is not applicable to non-gravitactic suspensions of \textit{CR} which have a  
spherical body. A sphere has a constant rotation velocity $\Omega=\dot{\gamma}/2$ in a simple shear flow and no anisotropy prevails. An estimation of the aspect ratio of a puller corresponding
to the experimentally
measured effective viscosity \cite{Peyla:2010.1} gives a value of ellipticity of about $7$, far above the value of $1$ for a spherical shape! 
Therefore, the origin of the viscosity enhancement for \textit{CR} suspensions remains an open question. 

Recently, Takatori and Brady \cite{Brady:2017.1} introduced a new idea and showed that the diffusion-like motion of micro-swimmers immersed in a simple shear flow, induces a non-zero
off-diagonal shear components in the swim stress tensor. This effect can explain the rheology of a suspension of {\textit CR}. 
In this work, we also point out a novel 
physical phenomenon that can also explain the peculiar rheology of \textit{CR} suspensions. 
We show that the flagella beating itself, leads to Jeffery-like orbits of the \textit{CR} in a simple shear flow even if a swimmer's rounded shape is considered. It is shown that
because of the coupling between the shear flow and the flagella beating, the swimmer can resist the rotation induced by the shear flow and thus, leads to an increase of effective viscosity. 
Somehow, our model helps to reconcile preceding theories done with rod-shape swimmers with spherical \textit{CR} rheology.

The use of a permanent force dipole to describe the swimmer activity \cite{Mehandia2008, Drescher2010,Peyla:2014.1}
is reasonable on long time scales. However, in this work, we show that even if the flow characteristic time  $\dot{\gamma}^{-1}$ 
is much larger than the beating period $T_b$, the flagellar breaststroke gives rise to peculiar Jeffery orbits of the swimmer and thus, affects the rheology of the dilute suspension
composed of these swimmers. 

In this work, we propose and investigate a three-bead model for a swimming \textit{Chlamydomonas Reinhardtii},
where the beads are connected by a frictionless scaffold. 
Our 3d model shares similarities with the two-dimensional 
models for \textit{CR},  described in Ref.~\cite{FriedrichB:2012.1} 
and in Ref.~\cite{Golestanian:2013.1}. 
While in both  works three equal-sized spheres are used, the beads representing the flagella are of different size 
in Ref.~\cite{Polotzek:2013.1}. The flagellar beads in these 2d models move on circular orbits. 
Here, we investigate within our 3d model the effect of the flagellar beating on a \textit{CR}
motility in a simple shear flow and its consequence on the rheology of a dilute suspension of \textit{CR}.
The model is presented in section $II$, the calculation of Einstein's viscosity
is given in section $III$ and our results are given in section $IV$.

\section{Model for  biflagellate algae}

\begin{figure}[hbt]
\begin{centering}
\includegraphics[scale=0.3]{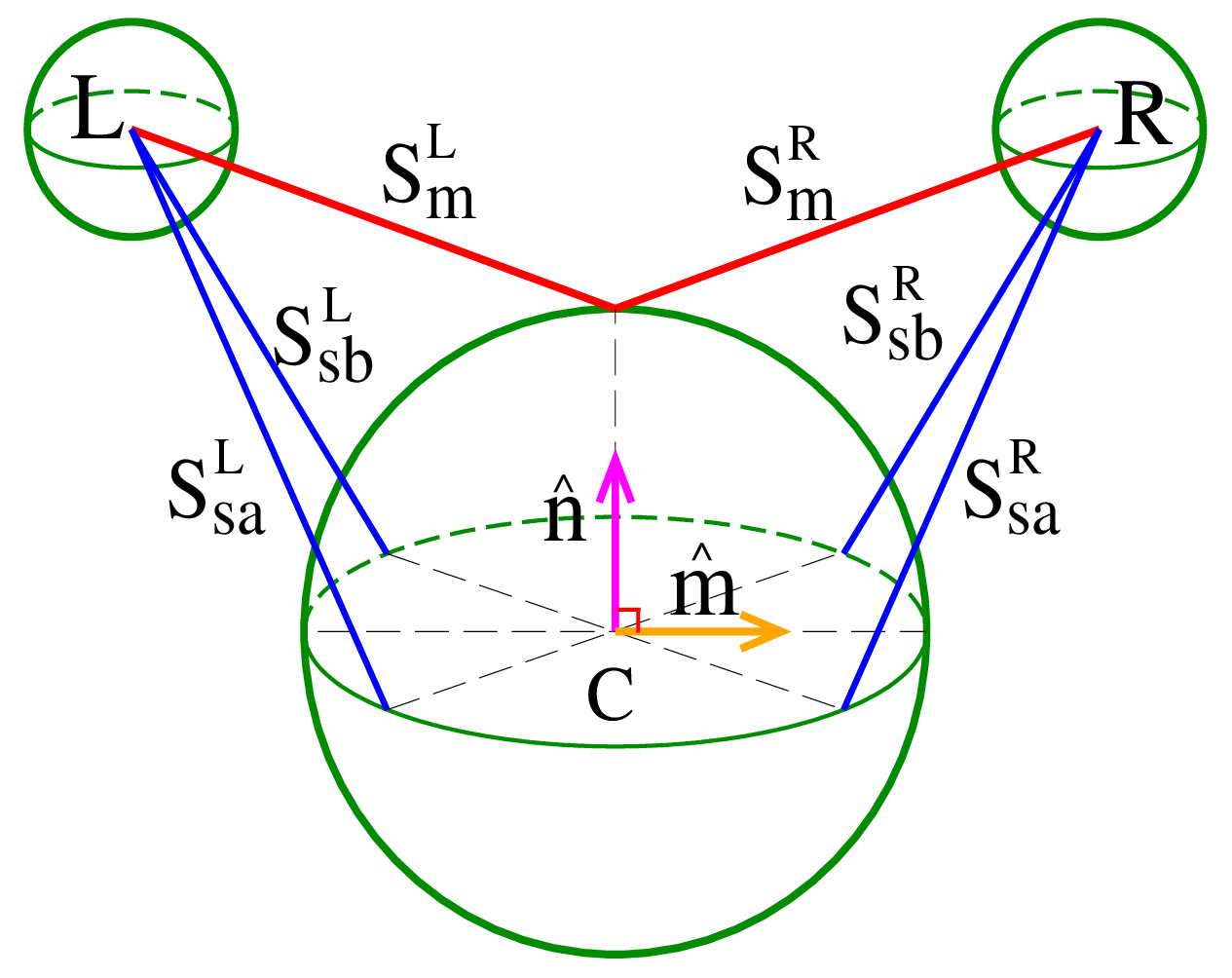} 

\par\end{centering}

\caption{\label{fig:Model} The active swimmer-model consists of three beads 
where the left ($L$) and the right ($R$) satellite   bead of radius $R'$
are connected to the body-bead of radius $R$  and centered in $C$
by frictionless  Hookean springs, $S_{sa,sb,m}^L$ or $S_{sa,sb,m}^R$,  respectively.
$R/R'=3$ and the equilibrium length of the springs can be chosen time-dependent or stationary. 
The unit vector $\mathbf{\hat{n}}$ defines the mean swimming 
direction and defines together with the unit vector $\mathbf{\hat{m}}$ 
the mean flagellar beating plane. }
\end{figure}

A three-dimensional swimmer-model  of a biflagellate alga \textit{Chlamydomonas Reinhardtii}
is described in this section. The body of the \textit{CR} alga is described by a sphere of radius $R$.
This is linked to two smaller satellite beads 
of radius $r$ mimicking the two flagella of the alga, as indicated
in  Fig.~\ref{fig:Model}. Each of the the left ($L$) and right ($R$) flagellar beads  is connected 
to the body by  three  springs $S_{m}^{L,\,R},~ S_{sa}^{L,\,R},~ S_{sb}^{L,\,R}$
(superscripts  $L$ or $R$ for left and right bead, respectively).
 The main springs $S_m^{L,R}$  are chosen more rigid 
than the two supporting springs  $S^{L,R}_{sa}$ and $S^{L,R}_{sb}$
(see table \ref{tab:Parameters}), which help to maintain 
both satellite beads into the  $(\hat {\bf m},\hat {\bf n})$ plane, cf. Fig.~\ref{fig:Model}.
This anchoring of each satellite bead also allows the application 
of  torques to the central bead, similarly to the flagella 
on the body of a \textit{CR}. In the case of an active swimmer-model,
the equilibrium lengths of the springs are time dependent, in order to 
mimic the characteristic breaststroke-like swimming motion of a \textit{CR}. For a dead swimmer equilibrium lengths 
become time-independent during a tumbling motion in a shear flow.

\begin{figure}[hbt]
\begin{centering}
\includegraphics[scale=0.3]{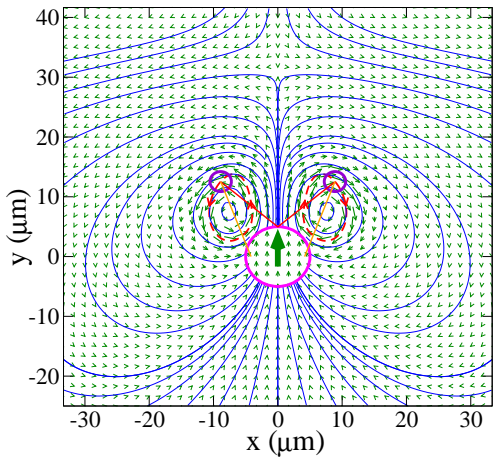}

\par\end{centering}

\caption{\label{fig:Model1}The velocity field and streamlines around the model swimmer in 
a fluid at rest (in the lab reference frame) and averaged over one period of the flagellar beating. The arrow inside the main sphere indicates the motion of the swimmer.
The closed dashed
 curves represent the orbits of each satellite bead with respect to the central bead. 
The swimmer moves back and forth during one period but in average it moves forward 
with a velocity $v\sim54 \, \mu m/s$ obtained for chosen parameters (see text and table \ref{tab:Parameters}). The flow lines as well as the velocity field compare very
well with the experimental results of Drescher \textit{et al.} \cite{Drescher2010}.}
\end{figure}

The two perpendicular unit vectors, $\mathbf{\hat{n}}$ and $\mathbf{\hat{m}}$, 
span the mean  ``flagella plane'' of the orbiting satellite beads.
 $\mathbf{\hat{n}}$ points from the center of the body to the common anchoring point
of the two main springs $S_{m}^{L,R}$ (Fig.~\ref{fig:Model})
and describes the mean swimming direction of a swimmer. The other unit vector
$\mathbf{\hat{m}}$  points from the center of the body to the midpoint of the
line between the two anchoring points of the supporting springs,  $S_{sa}^{R}$
and $S_{sb}^{R}$.  
The four anchoring points of the supporting springs  
build a square in the equatorial plane of the central bead with diagonals of length $2R$ and with  the
 vector $\mathbf{\hat{n}}$ normal to this square.

A swimming motion is induced when  the length
of  the Hookean springs, $\ell_i(t)$, 
is periodically modulated with a frequency $\omega=2\pi/T_b$: $\ell_i(t)=\ell_{i,0}+A_i\cos(\omega t+\varphi_i)$.
The spring length in the relaxed state is  $\ell_{i,0}$, the modulation amplitude  is $A_i$ and the phase $\varphi_i$.
Each satellite bead imposes via the springs, either  
during a swimming motion ($A_i\neq0$) or when the passive swimmer ($A_i=0$) tumbles in 
a shear flow,  via the springs  forces and  torques to the central bead.

The $25$ spring parameters $A_i$, $\ell_{i,0}$, $\varphi_i$ and the spring constant $k_i$ 
for each of the six springs as well as the frequency $\omega$ define the characteristics of the swimmer.
This number of parameters can be drastically reduced by assuming symmetries of the swimmer.
One is the mirror symmetry with respect to a plane perpendicular to $\mathbf{\hat{m}}$, {\it i.e.}
 parameters of
left and right springs are identical. We also assume that 
the swimmer is not rotating around 
its mean swimming direction $\mathbf{\hat{n}}$, \textit{i.e.} we can choose identical spring constants
for  $S_{sa}^{L,R}$ and $S_{sb}^{L,R}$.
Only the differences  between the phases of the oscillating main and 
supporting springs on each side are important, namely  
$\delta\varphi_{a,b}^{L,R}=\varphi_{sa,sb}^{L,R}-\varphi_{m}^{L,R}$
with $\delta\varphi_{a}^{L,R}=\delta\varphi_{b}^{L,R}=\delta \varphi$.
Hence, we are left with eight independent 
parameters: $\ell_{m0}$, $\ell_{s0}$, $A_{m}$, $A_{s}$, 
$\delta\varphi$, $\omega$, $k_{m}$ and $k_{s}$. 

The \textit{CR} alga has  approximately a spherical  shape with diameter $2R=10\mu m$
and for particles of this size the Brownian dynamics plays a minor role.
Considering the kinematic viscosity for water $\nu \approx 10^{-6}\,m^2 s^{-1}$ 
the Reynolds number is rather low: $Re =vR/\nu \sim 2.5\times10^{-4}$ for a typical swimmer velocity $v\approx 50 \mu m/s$. Therefore, inertial forces can be
neglected and the fluid flow can be described by
the Stokes equation  \cite{Brenner:1981}. The particle dynamics and 
hydrodynamic  particle-particle interactions can be described through a reflection method similar to the one used in the Rotne-Prager approximation \cite{Brenner:1981,Rotne:1969.1,Wajnryb:2013.1}.

Let us summarize our mathematical treatment below. Consider the $i^{th}$ sphere ($i=1..3$) with radius $R_i$.  The
particle is submitted to an external
force $\mathbf{F}_i$ and a torque $\mathbf{T}_i$ with no slip boundary condition. 
In our model, there
is no net force or net torque exerted on the swimmer. However, oscillating
springs can apply equal and opposite forces on beads that are aligned along the springs. 
Because of the design of our model, each spring
can exert a torque only on the central bead, whereby the total torque exerted  by the 
six springs on the central bead vanishes in a quiescent fluid.

At zeroth order, $\mathbf{v}_{0,i}=\mathbf{F}_i/(6\pi\eta R_i)$ and $\boldsymbol{\Omega}_{0,i}=\mathbf{T}_i/(8\pi\eta R_i)$ are 
respectively the translational velocity  and the rotational frequency  of the $i^{th}$ particle.
When the model swimmer in Fig.~\ref{fig:Model}) is placed in a quiescent fluid,
the flow field at zeroth order $\mathbf{u}_0$ expressed in $\mathbf{r}$
is:
\begin{equation}
\mathbf{u}_0(\mathbf{r})=\sum_{i=1}^{3} \mathbf{u}_{0,i}(\boldsymbol{\rho}_{i})=\sum_{i=1}^{3}\left [ \mathbf{V}^{t}_{0,i}(\boldsymbol{\rho}_{i})+\mathbf{V}^{r}_{0,i}(\boldsymbol{\rho}_{i})\right ],
\end{equation}
where $\boldsymbol{\rho}_{i}$ is the position vector pointing from the $i^{th}$ sphere's center to a given point $\mathbf{r}$ in the fluid: $\boldsymbol{\rho}_{i}=\mathbf{r}-\mathbf{r}_{i}$.
The summation is carried out on the three interacting spherical beads
of the model swimmer. The result 
averaged over one period of beating is represented in Fig.~\ref{fig:Model1} , \textit{i.e.} $2\pi/\omega$. 
Here, $\mathbf{V}^{t}_{0,i}$ and $\mathbf{V}^{r}_{0,i}$ are respectively the translational and
rotational parts of the flow field created by the moving and rotating $i^{th}$ sphere with
radius $R_i$, velocity $\mathbf{v}_{0,i}$ and angular velocity $\boldsymbol{\Omega}_{0,i}$ \cite{Landau6eng}:
\begin{eqnarray}
\nonumber
\mathbf{V}^{t}_{0,i}(\boldsymbol{\rho}_i)&=&\mathbf{v}_{0,i}\frac{3R_i}{4\rho_{i}}\left(1+\frac{R_i^{2}}{3\rho_{i}^{2}}\right)+ \\
&&+ \boldsymbol{\rho}_{i}(\mathbf{v}_{0,i}\cdot\boldsymbol{\rho}_{i})\frac{3R_i}{4\rho_{i}^{3}}\left(1-\frac{R_i^{2}}{\rho_{i}^{2}}\right),
\label{eq:Stokes flow} \\
\mathbf{V}^{r}_{0,i}(\boldsymbol{\rho}_i)&=&\left(\frac{R_i}{\rho_{i}}\right)^{3}\boldsymbol{\Omega}_{0,i} \times \boldsymbol{\rho}_{i}.
\end{eqnarray}
This, in turn, influences the motion of the $j^{th}$ ($j=1..3$) particle centered in $r_j$
which is calculated via the Fax\'en laws \cite{Brenner:1981} at first order:
\begin{eqnarray}
\mathbf{v}_{j,1}&=&\frac{\mathbf{F}_j}{6\pi\eta R_j}+\sum_{i \neq j}^{3}(1
+\frac{R_j^{2}}{6}\triangle)\mathbf{u}_{0,j}|_{r_{j}},\label{eq:transl}\\
\boldsymbol{\Omega}_{j,1}&=&\frac{\mathbf{T}_j}{8\pi\eta R_j}+\frac{1}{2} \sum_{i \neq j}^{3}\boldsymbol{\nabla}\times\mathbf{u}_{0,j}|_{r_{j}}\,.\label{eq:rot}
\end{eqnarray} 
The quantities $\mathbf{u}_{0,i}$, $\triangle \mathbf{u}_{0,i}$ and $\boldsymbol{\nabla}\times\mathbf{u}_{0,i}$ 
are estimated at the center
of the $j^{th}$ particle.
When the model swimmer is immersed in a shear flow \cite{mikulencak2004}, the shear induced velocity
$\mathbf{V}^{*}_{i}(\boldsymbol{\rho}_i)$ must be added to the flow field $\mathbf{u}_{0,i}(\boldsymbol{\rho}_i)$ 
for each spherical
bead $i$: 
\begin{eqnarray}
\nonumber
V_{x,i}^{*} (\boldsymbol{\rho}_i) &=&  \dot{\gamma}\left [ y_i-\frac{y_i R_i^5}{2\rho_i^{5}}-\frac{5}{2}\frac{x_i^{2}y_i}{R_i^2}\left(\frac{R_i^{5}}{\rho_i^{5}}-\frac{R_i^{7}}{\rho_i^{7}}\right)\right ] \,,\\
\nonumber
V_{y,i}^{*} (\boldsymbol{\rho}_i) &=&  \dot{\gamma}\left [ -\frac{x_i R_i^5}{2\rho_i^{5}}-\frac{5}{2}\frac{x_iy_i^{2}}{R_i^2}\left(\frac{R_i^{5}}{\rho_i^{5}}-\frac{R_i^{7}}{\rho_i^{7}}\right)\right ] \,,\\
V_{z,i}^{*} (\boldsymbol{\rho}_i) &=&  -\dot{\gamma}\frac{5}{2}\frac{x_iy_iz_i}{R_i^2}\left(\frac{R_i^{5}}{\rho_i^{5}}-\frac{R_i^{7}}{\rho_i^{7}}\right)\,.
\label{shearinducedvelocity}
\end{eqnarray}
Here $\rho_i=\sqrt{x_i^{2}+y_i^{2}+z_i^{2}}$, $xOy$ is the shear plane and
$\dot{\gamma}$ is the imposed shear rate.
Using the approximation of the Fax\'en theorems and keeping the first iteration
step in the reflection of hydrodynamic interactions leads to a 
method valid up to order of $(R/r)^{3}$ \cite{Reichert2006}.

For a given set of parameters,
including the
oscillation amplitudes $A_m$,  the two 
satellites move in general on elliptical orbits
as indicated in Fig.~\ref{fig:Model1}. 
The left (resp. right) bead is orbiting counter-clockwise (resp. clockwise). This 
bead motion mimics the motion of the flagella of \textit{CR} \cite{Peyla:2011.1}: The
model swimmer moves forward (along $\mathbf{\hat{n}}$) during a power stroke (\textit{i.e.} when satellite beads move 
in the direction of the body)
and moves backward during a recovery stroke (\textit{i.e.} when satellite beads move in the opposite direction).
This resembles very much the experimentally observed vacillating swimming motion of \textit{CR}
\cite{Peyla:2011.1}. 
Fig.~\ref{fig:Model1}
 shows besides the orbits of the satellite beads  the velocity 
field and the streamlines averaged over one period of the motion of the satellite beads.

The parameters in table \ref{tab:Parameters} are chosen in such a way that they
reproduce rather closely the swimming 
characteristics of a \textit{CR}. For instance, the swimming velocity of the model 
(averaged over the orbiting period of the satellite beads 
$T_{b}=2\pi/\omega=1/50s$) is  
$v=540 R \omega \sim 54\mu m/s $ considering a radius $R= 5 \mu m$. 
Note that if spring parameters are chosen such  that 
$k_{m}/k_{s}<1$ and $\delta\varphi\sim\pi$, 
then the motion of the satellite beads along their  orbits is reversed, and the swimmer 
would move 
backward along the  $-\mathbf{\hat{n}}$ direction like a pusher. 

\begin{table}[hbt]
\caption{\label{tab:Parameters} Parameter values of our model.}

\centering{}%
\begin{tabular}{|>{\centering}p{1.0cm}||>{\centering}p{0.4cm}|>{\centering}p{0.6cm}|>{\centering}p{0.7cm}|>{\centering}p{0.7cm}|>{\centering}p{0.6cm}|>{\centering}p{0.8cm}|>{\centering}p{0.8cm}|>{\centering}p{0.6cm}|}
\hline 
 & {\large $\frac{R}{r}$} & {\large $\frac{k_{s}}{k_{m}}$} & {\large $\frac{\ell_{m0}}{R}$} & {\large $\frac{\ell_{s0}}{R}$} & {\large $\frac{A_{m}}{R}$} & {\large $\frac{A_{s}}{R}$} & {\large $\frac{\omega}{2\pi}$} & $\delta\varphi$\tabularnewline
\hline 
\hline 
{\footnotesize active} & $3$ & $4/7$ & $1.9$ & $1.8$ & $1.3$ & $0.8$ & $50Hz$ & $\pi/3$\tabularnewline
\hline 
{\footnotesize inactive} & $3$ & $0$ & $1.9$ & - & - & - & - & -\tabularnewline
\hline 
\end{tabular}
\end{table}

For any asymmetry 
between the spring parameters of the left and right satellite bead,  swimming becomes 
less efficient compared to the case of a symmetric and synchronous motion  of the satellite beads.
Moreover, the mean swimming direction  is not stationary anymore and moves for instance on curved trajectories  
or leads to other complex swimming
trajectories as  reported
for \textit{CR} \cite{sourcebook}.  

The model microswimmer with oscillating harmonic springs  imitates the swimming of a \textit{CR} 
and is referred to as ``active swimmer''. We call it ``inactive swimmer'' or dead swimmer  when the 
supporting springs $S_{sa}^{L,R}$ and $S_{sb}^{L,R}$ are removed and the modulation amplitudes  $A_{m}$ vanish. 
Then, the satellite beads   are very flexible connected  to the body, i. e. they are allowed to move freely in the vicinity of the central bead 
due to an external flow, similar to the flagella of a dead \textit{CR}.

\section{Intrinsic viscosity of a dilute suspension of microswimmers.}

\begin{figure}[hbt]
\ \ \includegraphics[scale=0.43]{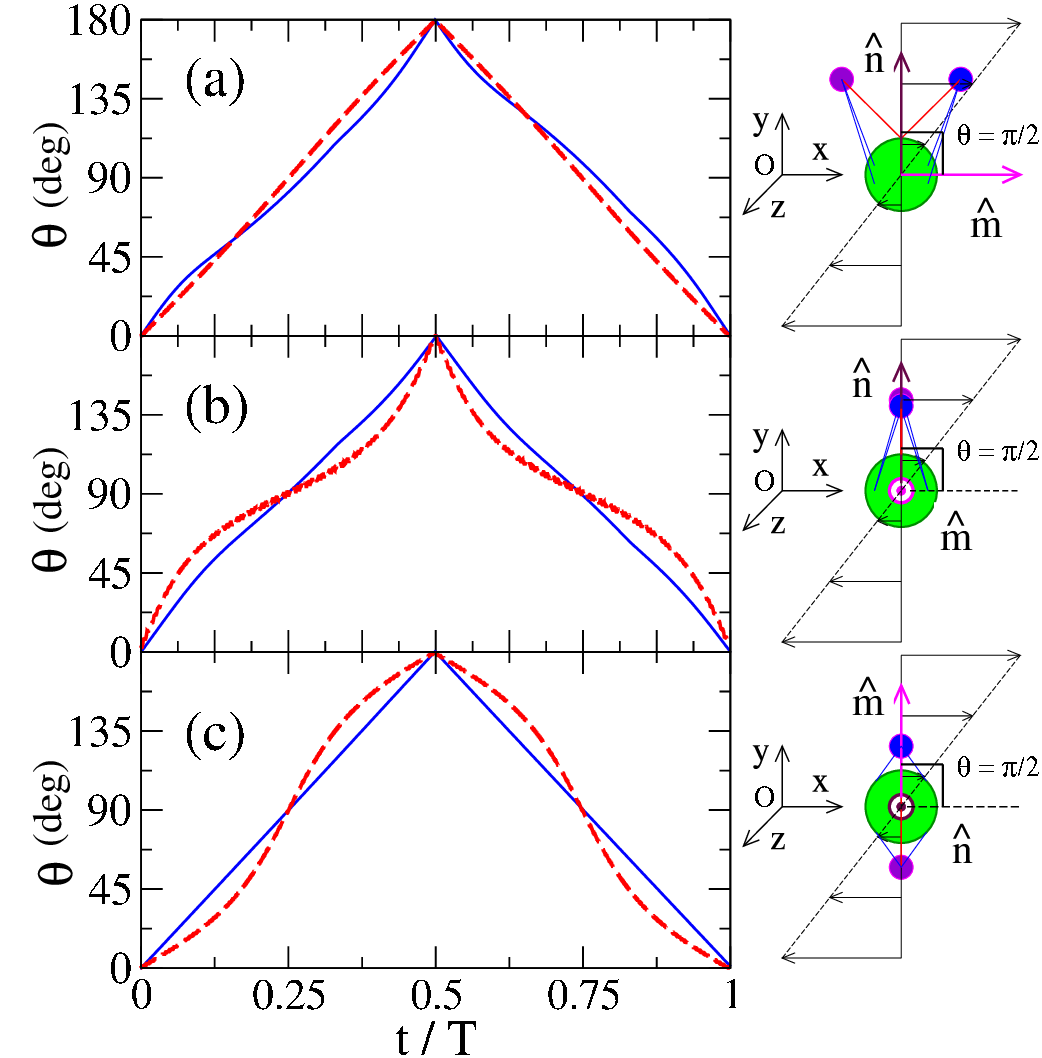}

\caption{\label{fig:orbits}Temporal orientation angles  of  active (dashed lines)  or inactive microswimmers (solid line)
in a linear shear flow for three different swimmer orientations.
 (a) $\mathbf{\hat{n}}$ (swimming direction) and $\mathbf{\hat{m}}$ are both parallel to the shear ($xy$) plane. 
 (b) $\mathbf{\hat{n}} $ parallel to the  $xy$ plane and  $\mathbf{\hat{m}} \parallel \hat {\bf z}$. 
 (c) $\mathbf{\hat{m}} \parallel$  to  the shear plane and $\mathbf{\hat{n}} \parallel \hat {\bf z}$. 
 In parts (a) and (b) $\theta$ is the angle between $\mathbf{\hat{n}}$ and  $ \hat {\bf x}$, while in part
 (c) $\theta$ is the angle between $\mathbf{\hat{m}}$ 
 and  $\hat {\bf x}$. The time dependence is plotted  in units of the shear rate dependent tumbling period $T$,
 which is different for each curve [see table \ref{tab:tumbling_period}].}
\end{figure}

In this section, we consider a  single microswimmer in a linear shear flow $\mathbf{v}=(\dot{\gamma}y,\,0,\,0)$
with the shear rate $\dot{\gamma}$ and its contribution to  the effective shear viscosity. Hydrodynamic
interactions among swimmers are neglected, which  corresponds to a dilute swimmer suspension.

For a dilute suspension of rigid particles the effective viscosity depends linearly on the volume fraction $\phi$
occupied by the particles:
\begin{equation}
\eta_{eff}=\eta(1+\alpha\phi).
\label{effectiveviscosity}
\end{equation}
$\eta$ is the shear viscosity of the solvent and the  dimensionless 
coefficient $\alpha$ is the intrinsic viscosity and $\phi=V_0/V$ where $V_0=4/3 \pi (R^3+2R'^3)$ is the volume of the swimmer and $V$ is the total volume of the suspension (swimmer and water). 
The quantity $\alpha$ is known as the
Einstein\textquoteright{}s intrinsic viscosity \cite{Einstein06,Einstein11}. In general, $\alpha$ is defined as 
\begin{equation}
\alpha=\lim_{\phi\rightarrow0}\frac{\eta_{eff}-\eta}{\eta\phi}.
\end{equation}
 Following the Batchelor method and neglecting inertia, the 
average stresslet $S^{(p)}_{ik}$ induced by a  particle in the fluid is as follows  \cite{Batchelor70,Landau6eng}:
\begin{equation}
S^{(p)}_{ik}=\oint_\mathcal{A}\left\{ \sigma_{ij}x_{k}\mathcal{N}_{j}-\eta(v_{i}\mathcal{N}_{k}
+v_{k}\mathcal{N}_{i})\right\} dA\label{eq:sigma_p}\,.
\end{equation}
$\boldsymbol{\hat{\mathcal{N}}}$ is the unit outward vector normal to the surface $\mathcal{A}$ encompassing a volume $V$. 
Any closed surface $\mathcal{A}$ can be selected for integration as long as 
it contains  the swimmer. The average total shear stress in a volume $V$ containing a single swimmer is $\Sigma_{xy}=\eta {\dot \gamma}+S^{(p)}_{xy}/V=\eta {\dot \gamma}+S^{(p)}_{xy}\phi/V_0$, $V_0$ being 
the volume of the swimmer. 
Note that in the absence of an external torque imposed on the swimmer $\boldsymbol{S}^{(p)}$ is symmetric
and  only the deviatoric part $\Sigma_{xy}$ of the stress is significant \cite{Batchelor70}. 
The effective viscosity is then calculated such as $\eta_{eff}=\Sigma_{xy}/{\dot \gamma}$ 
which gives via Eq.~(\ref{effectiveviscosity}):
\begin{equation}
\alpha=\frac{1}{\eta \dot{\gamma} V_0}S^{(p)}_{xy}\label{eq:alpha}\,.
\end{equation}
In the case of a single rigid sphere the exact velocity 
and pressure fields around the sphere are known analytically and the calculation of $\alpha$ 
gives the well known value: $\alpha=5/2$ \cite{Einstein06,Einstein11}.
\begin{table}
\caption{\label{tab:intrinsic_viscosity}Intrinsic viscosity for different swimmer orientations. }

\begin{tabular}{|>{\centering}p{3.0cm}||>{\centering}p{1.8cm}|>{\centering}p{1.8cm}|}
\hline 
 & $\alpha$ 

inactive & $\alpha$ 

active\tabularnewline
\hline 
\hline 
$\mathbf{\hat{n}}\,\angle\, xy;$ $\mathbf{\hat{m}}\,\angle\, xy$ & $3.0$ & $5.7$\tabularnewline
\hline 
$\mathbf{\hat{n}}\,\angle\, xy;$ $\mathbf{\hat{m}}\parallel z$ & $2.9$ & $4.8$\tabularnewline
\hline 
$\mathbf{\hat{m}}\,\angle\, xy;$ $\mathbf{\hat{n}}\parallel z$ & $2.5$ & $3.5$\tabularnewline
\hline 
all configurations & $2.7$ & $4.4$\tabularnewline
\hline 
experimental \cite{Peyla:2010.1,Mussler:2013.1} & $2.5\pm0.1$ & $4.5\pm0.2$\tabularnewline
\hline 
\end{tabular}

\caption{\label{tab:tumbling_period}Period of  swimmer tumbling. 
Note that the tumbling period for a single sphere depends on the shear rate $\dot  \gamma$:
$T=2\pi/\left|\Omega\right|=4\pi/\dot{\gamma}\simeq126T_{b}$.}

\begin{tabular}{|>{\centering}p{3.0cm}||>{\centering}p{1.8cm}|>{\centering}p{1.8cm}|}
\hline 
 & $T/T_{b}$ 

inactive & $T/T_{b}$

active \tabularnewline
\hline 
\hline 
$\mathbf{\hat{n}}\,\angle\, xy;$ $\mathbf{\hat{m}}\,\angle\, xy$ & $126$ & $146$\tabularnewline
\hline 
$\mathbf{\hat{n}}\,\angle\, xy;$ $\mathbf{\hat{m}}\parallel z$ & $145$ & $166$\tabularnewline
\hline 
$\mathbf{\hat{m}}\,\angle\, xy;$ $\mathbf{\hat{n}}\parallel z$ & $127$ & $150$\tabularnewline
\hline 
\end{tabular}
\end{table}

A significant enhancement of the effective viscosity for a suspension of living \textit{CR}  is reported from experiments  \cite{Peyla:2010.1,Mussler:2013.1}.
Hydrodynamic interactions among swimmers as well as confinement effects \cite{Sangani:2013.1}
may play an important role for concentrated suspensions. 
Here, our analysis based on an isolated single swimmer does not include such effects and is limited to dilute suspensions ($\phi<0.01$),
which is usually the case for a natural planktonic suspension. 
We compare the numerically calculated intrinsic viscosity, $\alpha$, for dilute suspensions to the experimentally observed  
value $\alpha_{exp}$ \cite{Peyla:2010.1}, estimated by
a fit  according to the Krieger and Dougherty's law \cite{Krieger59}.

The temporal modulation of the equilibrium length of the springs 
and the exposition of the swimmer to the linear shear flow leads to a complex dynamics of the three beads
and therefore to  a time-dependent intrinsic viscosity $\alpha$. The flow field around the swimmer is a superposition of the
imposed linear shear flow and flow perturbations caused by the bead dynamics. Then the numerical surface integration in Eq.~(\ref{eq:sigma_p}) provides  a value for $\alpha(t)$ [through Eq.~(\ref{eq:alpha})]
for each temporal configuration of the microswimmer, {\it i.e.} at each time $t$.

The calculation of the intrinsic viscosity $\alpha(t)$ at  a chosen time $t$ via Eq.~(\ref{eq:alpha}) 
strongly depends on the relative positions of satellite beads on their orbits. 
It also depends on the orientation of the microswimmer  and of its flagella plane $(\mathbf{\hat{n}},\mathbf{\hat{m}})$ 
with respect to the flow direction ($\hat {\bf x}$) and the shear is the plane ($xy-$plane). 
Therefore, $\alpha$ is averaged over the tumbling period $T$ of the microswimmer 
in a shear flow (see table \ref{tab:intrinsic_viscosity}). Note  
that one has usually $T \gg T_b=2\pi/\omega$ (c.f. table \ref{tab:tumbling_period}), 
where $T$ is of the same order of magnitude 
than the period of the rotational part of the  shear flow: $2 \pi/\Lambda=4\pi/\dot{\gamma}$ (with $\Lambda \sim \dot{\gamma}/2$). 
The tumbling period of swimmers depends also on the swimmer's orientations 
[see Table \ref{tab:tumbling_period}]. In dilute suspensions, we assume that any  orientation of a swimmer
is equally probable, which  may not be true for gravitactic microswimmers - not considered here. 
Therefore, before comparing our numerical results for $\alpha$
with experimental measurements, we take the ensemble average of different realizations of the 
swimmer's orientations with respect to the flow direction and the shear plane.
Results are discussed in the following section.

\section{Results and discussion}

The  intrinsic viscosity of the model microswimmer is evaluated
by averaging over all swimmer orientations and
for the shear rate $\dot{\gamma}=5s^{-1}$, which corresponds to the experimental value 
\cite{Peyla:2010.1,Mussler:2013.1}.
For an active swimmer suspension, we obtain $\alpha=4.4$ and for an inactive swimmer suspension $\alpha=2.7$. The intrinsic viscosity,
of the active swimmer suspension is consistent with the experimentally measured viscosity $\alpha_{exp}=4.5\pm0.2$ in
Ref.  \cite{Peyla:2010.1} and  its $\alpha_{exp}=4.5\pm0.17$ in Ref.~\cite{Mussler:2013.1}. 
For the suspension of inactive swimmers 
the numerically calculated intrinsic viscosity is also comparable to the experimental value $\alpha_{exp}=2.5\pm0.1$
\cite{Peyla:2010.1}.

In our averaging, all orientations of the swimmer in the shear flow are possible and each of them corresponds 
to a different Jeffery orbit. Next, we consider some specific orientations of active and inactive swimmers 
with respect to the shear flow, for exploring the origin of the viscosity enhancement. 
In the first configuration in Fig.~\ref{fig:orbits}(a) swimmers  tumble  
with $\hat {\bf m} $ and $\hat {\bf n}$   in the shear plane ($xy$). In this case, the intrinsic viscosity for an
active swimmer is $\alpha=5.7$ [see table \ref{tab:intrinsic_viscosity}] and for
an inactive swimmer we obtain a smaller value  $\alpha=3.0$. 
The angular velocity  of an active swimmer is surprisingly uniform, as indicated by the dashed curve in Fig.~\ref{fig:orbits}(a).
In addition, the tumbling period $T\simeq146T_{b}$ of an active swimmer is considerably  larger than for an  inactive one, $T\simeq126T_{b}$,
which is equal to the  tumbling period of a sphere of the same radius,  $T=2\pi/\left|\Lambda\right|=4\pi/\dot{\gamma}\simeq126T_{b}$.

A more unexpected orbit has been found for the active swimmer with $\mathbf{\hat{n}}$  
in the shear plane and $\mathbf{\hat{m}}$ parallel to $\hat {\bf z}$, cf. Fig.~\ref{fig:orbits}(b).
The active swimmer rotates slower when  $\mathbf{\hat{n}}$ is roughly parallel to $\pm y$ ($\theta\sim\pi/2$) 
and rotates faster with $\mathbf{\hat{n}}$ nearly parallel to the flow, \textit{i.e.} $\theta\sim0$ or $\theta\sim\pi$
[see dashed curve on Fig. \ref{fig:orbits}(b)]. This is in contrast to the behavior of  
an elongated object in a shear flow. For the configuration in Fig.~\ref{fig:orbits}(b)
the tumbling period and the intrinsic viscosity for inactive swimmer,  $T\simeq145T_{b}$ and $\alpha=2.9$, are
again considerably smaller than for an active swimmer with $T\simeq166T_{b}$ and $\alpha=4.8$.

\begin{figure}
\includegraphics[scale=0.32]{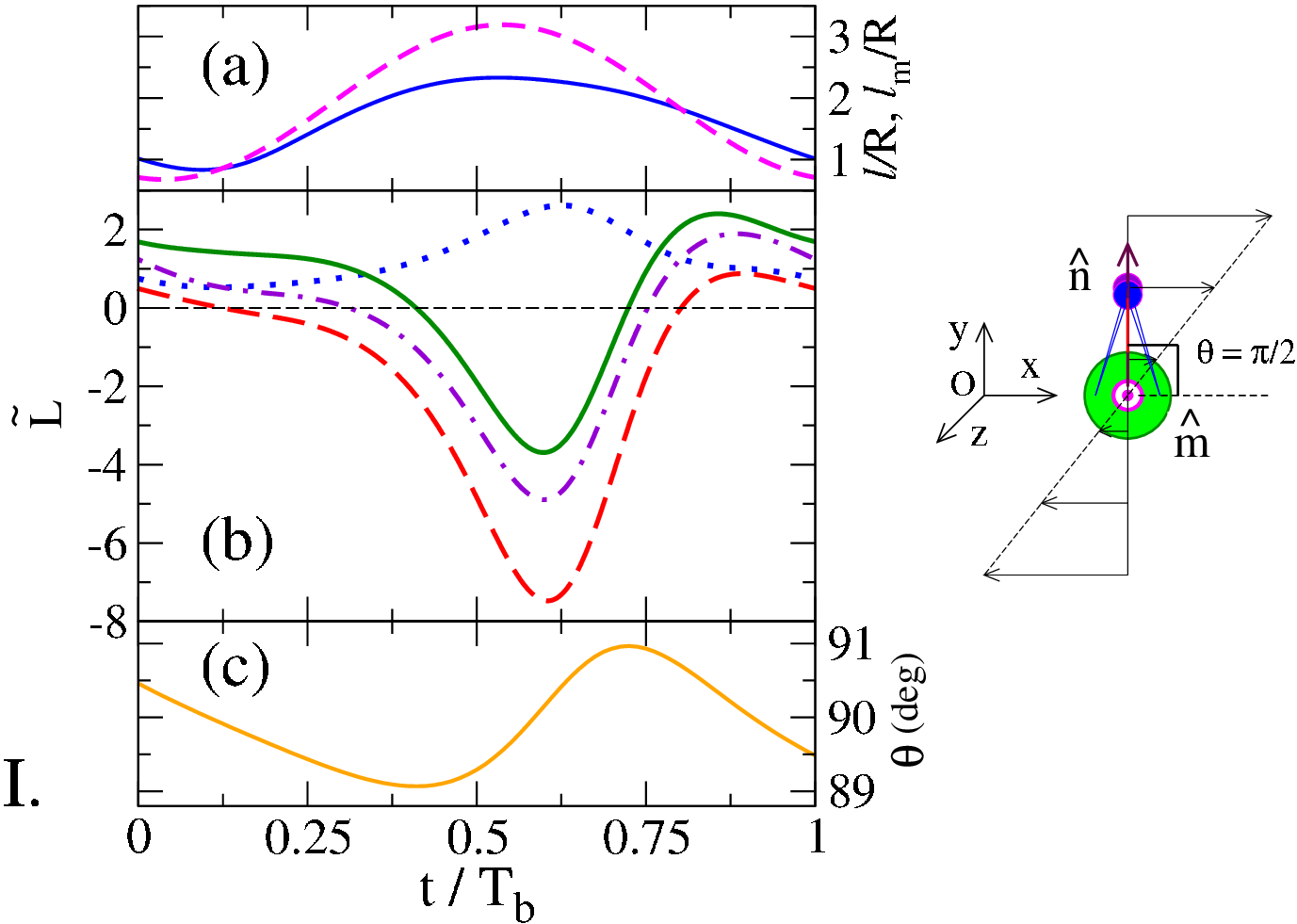}

\includegraphics[scale=0.32]{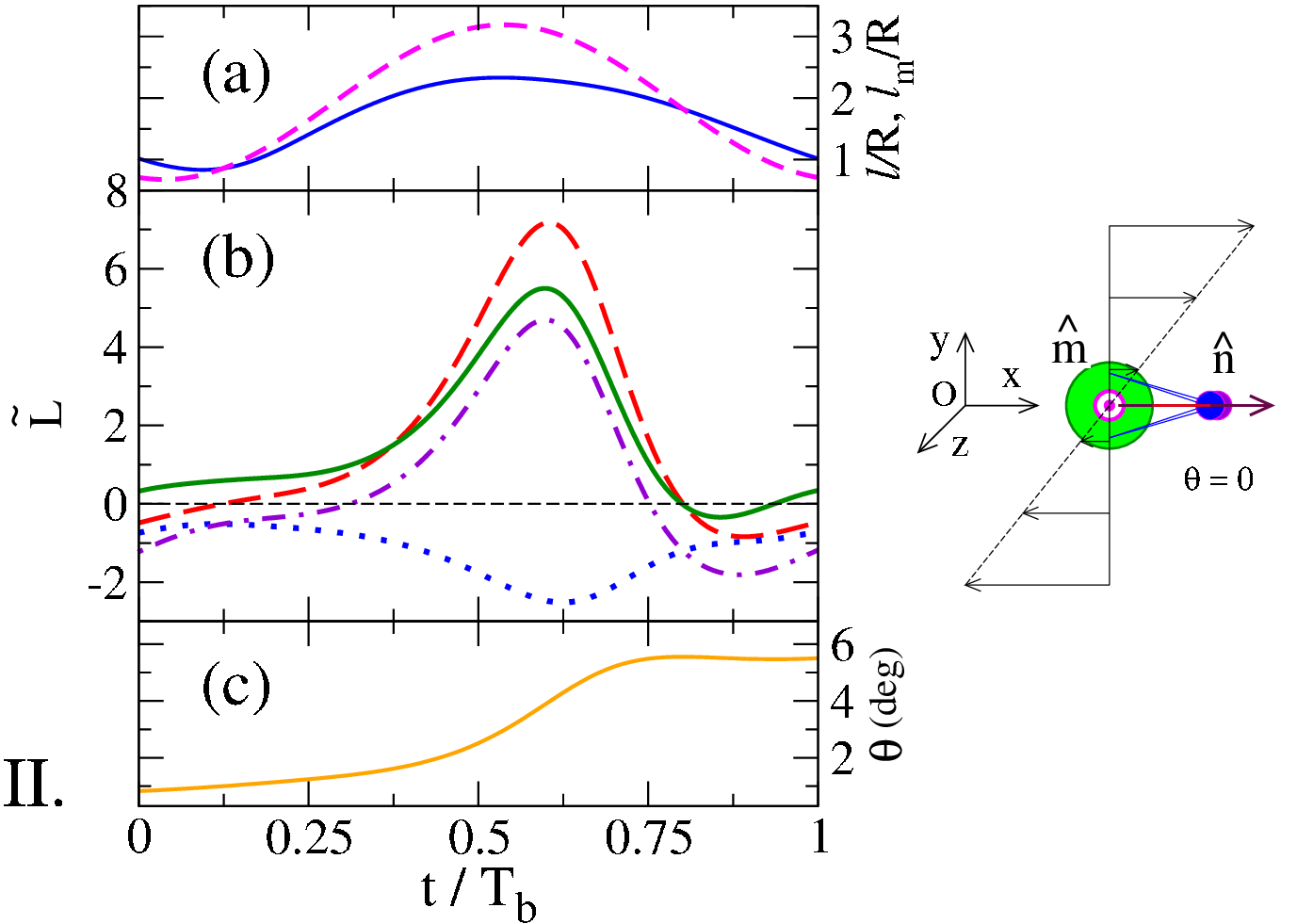}
\caption{\label{fig:ang_vel}
The top part {\bf I} is for the swimming direction perpendicular to the flow direction, i. e. 
$\mathbf{\hat{n}}\parallel y$, and $\mathbf{\hat{m}}\parallel z$. The bottom part {\bf II} is for the swimming direction  parallel 
to the flow direction, i. e. $\mathbf{\hat{n}}\parallel x$, and $\mathbf{\hat{m}}\parallel z$. In {\bf I} and in {\bf II} part (a) 
shows the time-dependence of the actual length $l(t)$ of the main springs in units of $R$ (solid curve) and the 
time-dependent equilibrium length $l_m(t)$ of the main springs (dashed curve) during
one period $T_b$ of the satellite bead motion. Part (b) shows in  {\bf I} and in {\bf II}
the 
dimensionless torque originating from the main springs (dashed curve), the supporting springs (dotted curve) and all 
spring together (dashed-dotted curve). Solid curve shows the total torque exerted by the satellite beads on the body. 
(c) shows the orientation angle $\theta(t)$ of the swimmer in the shear flow. 
}
\end{figure}

We find the same trend for the third configuration in Fig.~\ref{fig:orbits}(c), where we consider 
a swimmer where $\mathbf{\hat{m}}$ is in the shear plane and $\mathbf{\hat{n}}$ parallel to  $\hat {\bf z}$.
Here, the active swimmer is tumbling 
on usual Jeffery-like orbits [see Fig. \ref{fig:orbits}(c)]. 
Again the  tumbling period and the intrinsic viscosity of an inactive swimmer are respectively  $T\simeq127T_{b}$ and $\alpha=2.5$, and
are  smaller than for a active swimmer:  $T\simeq150T_{b}$ and $\alpha=3.5$.

In the case of any other arbitrary orientation of a swimmer with respect to the shear plane, the effective viscosity smoothly 
changes between the corresponding configurations discussed before. The enhancement of the intrinsic viscosity
and the reduction of the angular velocity for active swimmers in comparison to inactive swimmers have their origin 
on the one hand 
in the different torques active and passive swimmers experience during their tumbling dynamics in a shear flow.
On the other hand, the satellites are kept by the springs at a  certain mean distance from the main body,
which enhances the effective diameter of the active swimmer compared to the inactive swimmer. Indeed, for an inactive swimmer,  
satellites are fixed only by one  spring and  are therefore very flexible like the flagella of a dead \textit{CR}.

The satellite beads anchored to the body by  springs with time-dependent equilibrium lengths in combination with the shear flow 
exert a torque on the cell body. This causes  changes of the angular velocity of an active swimmer, compared 
to the one of a rigid sphere without satellites.
When a swimmer with mirror symmetry swims without an external flow, each spring exerts 
a torque on the body, but the sum of the torques exerted by all springs is zero and swimmer moves along spontaneous (initial) 
orientations. 
In Figure \ref{fig:ang_vel}, two configurations are considered with the flagella plane  perpendicular to the shear plane, 
and the swimming direction either
 perpendicular (I) or parallel (II) to the flow direction.
 We define the dimensionless torque,  $\widetilde{L}=-L/(4\pi\eta R^{3}\dot{\gamma})$, 
 imposed by the flagella on the swimmer's body.
$L$ is the torque applied on the body by the satellites via springs and $L_s=-4\pi\eta R^{3}\dot{\gamma}$ 
is the torque imposed by the shear flow on a spherical body of the same radius but without satellite beads. 
For $\widetilde{L}<0$  the torque experienced by the swimmer is opposite to the torque exerted by the 
shear flow.

\begin{figure}
\includegraphics[scale=0.28]{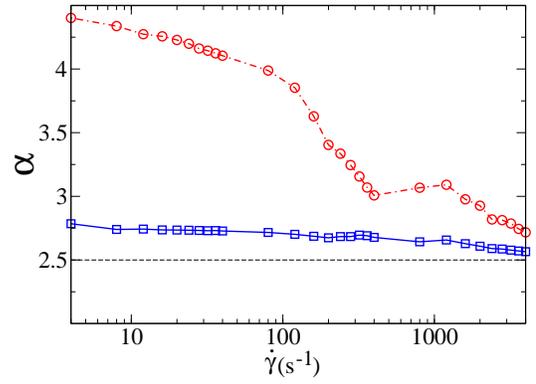}

\caption{\label{fig:shear_thinning}The intrinsic viscosity as a function of the shear rate. 
Solid line (squares) corresponds to  inactive swimmers and the 
dashed line (circles)  to the active swimmers.}
\end{figure}

Figure \ref{fig:ang_vel}-I shows the swimmer dynamics when flagella plane and swimming direction 
are both perpendicular to the flow direction. 
Figure \ref{fig:ang_vel}~I(a) shows the time dependence of the  lengths of the main springs, $\ell(t)$ (solid curve), 
and the imposed time-dependence of the equilibrium length  of the main springs, $\ell_{m}(t)$ (dashed curve), 
during an orbital period $T_b$ of the satellite beads. When the main springs 
are compressed ($\ell<\ell_{m}$), the torque exerted by the main springs to the body is 
opposite to the torque applied by the shear flow. 
Meanwhile, the supporting springs exerts a torque on the body which have the same sign as the torque applied by the shear flow. 
However, the total torque exerted by all springs on the body is opposite to the one imposed by the shear flow. 
Altogether,  the angular velocity of the swimmer is reduced, and the swimmer stays longer along 
the swimming direction perpendicular to the flow, in  comparison to a spherical particle of the same radius as
the body bead Fig.~\ref{fig:orbits}~I(b) [see also table \ref{tab:tumbling_period}]. 
It is remarkable that such a behavior is opposite to the usual Jeffery orbit of a passive elongated object in a shear flow 
\cite{JefferyGB:1922.1}. 

In contrast, when the flagella plane is still perpendicular to the flow but with the swimming direction parallel to the flow 
as in  Fig. \ref{fig:ang_vel}-II, 
the swimmer behaves in an opposite way. 
The total torque exerted by all springs on the 
swimmer amplifies the rotation imposed by the flow when the main spring is compressed ($\ell<\ell_m$). 
Again, the opposite effect is observed for an elongated particle aligned in the shear flow direction.
The leading effect, that slows down or accelerates the angular velocity, comes from the main 
springs to the satellite beads (flagella)  and the supporting springs  reduce the effect. 

When the flagella plane and the shear plane are identical, each main spring exerts a constant torque during the tumbling rotation of the 
swimmer such that the angular velocity is almost constant on the orbit as in  Fig. \ref{fig:orbits}(a).

We calculated the intrinsic viscosity over a wide range of shear rates $\dot \gamma$
and found a {\it shear thinning} behavior as shown in Fig.~\ref{fig:shear_thinning}. This 
is very similar to the experimentally observed shear thinning  \cite{Peyla:2010.1}.
The shear thinning for an
active microswimmer suspension is related to the existence of two timescales: 
the orbiting period $T_{b}$ of the satellite beads and the tumbling period $T$ depending on the shear rate. 
In the range $T/T_{b}>100$ (\textit{i.e.} $\dot{\gamma} \lesssim \omega/50$) one has many breast-strokes per tumbling period $T$.
Therefore, the satellite motion
can significantly influence the swimmers orientation thus enhancing the swimmers tumbling period.
Increasing the  shear rate decreases the tumbling period, the number of breast-strokes per tumbling period 
decreases as well as the influence of the swimming motion on the swimmers orientation. We observe a small enhancement of the intrinsic viscosity around $400-1000Hz$. 
This is a model related effect. At a shear rate of $5Hz$, the ratio $T/T_b$ is about $150$ (see table \ref{tab:tumbling_period}). 
In the range of shear rates, $400-1000Hz$,  the time scales $T$ and $T_b$ become comparable.
The re-increase of the viscosity is due to a synchronization between 
the rotating beads and the tumbling and depends on the parameters of the model. This effect  is probably absent for a real \textit{CR}.

Note that the shear thinning behavior is almost vanishing for an inactive swimmer, because the satellite beads are only connected by one
spring and therefore very flexible. 
When each satellite bead is still  connected by three springs but the equilibrium length are kept constant (inactive), then 
the effective radius of the swimmer is enhanced compared to the inactive swimmer. A  configuration of rather rigidly connected
satellite beads  does not correspond to the  flexible 
flagella of a dead \textit{CR}. However, 
this extension  of the effective radius contributes besides the
active swimming motion also to the enhanced intrinsic viscosity. However, this contribution 
reduces when increasing the shear rate due to the strong deformations of springs.

\section{Conclusion}

A three-dimensional bead spring model has been developed for \textit{Chlamydomonas Reinhardtii} that takes into account 
the  flagellar beating. 
The model correctly reproduces most of the swimming characteristics of this microswimmer. 
Using the model, we found a reversed Jeffery 
like orbit for the microswimmer in the shear flow. 
Such an altered orbit is essentially at the origin of the enhancement of suspension viscosity.
We determine the intrinsic viscosity of an active  and an inactive swimmer suspension using numerical  simulations of the Stokesian dynamics
of the three bead model within the 
generalized Rotne-Prager approximation. Our numerical results for the intrinsic viscosity are very similar 
to previous experimental measurements 
\cite{Peyla:2010.1,Mussler:2013.1} 
including a shear thinning behavior.

Our numerical results suggest that the significant increase of the viscosity for puller-type active microswimmer suspensions can be explained by considering the activity of 
an individual swimmer without a collective behavior. This effect could probably be combined with the diffusive trajectory of a  \textit{CR} which also affects the effective viscosity 
\cite{Brady:2017.1}
of dilute suspensions.

The complex angular orbits of the model swimmer and its consequences on the suspension viscosity for different orientations of the swimmer with respect of the 
flow direction, emphasize the importance of using a three-dimensional model for such a system.

{\it {Acknowledgments.-}}
We all acknowledge the French-German University (AUF) and the French-German Doctoral School "Living Fluids".

\end{document}